\documentstyle[preprint,aps]{revtex}
\begin{document}
\draft

\preprint{WU-AP/51/96, astro-ph/9601056}
\title{Fluctuations of the gravitational constant\\
induced by primordial bubbles}
\author{Nobuyuki Sakai\thanks{Electronic address: sakai@cfi.waseda.ac.jp}}
\address{Department of Physics, Waseda University, Shinjuku-ku, Tokyo 169,
Japan}
\date{Revised April 10 1996}
\maketitle

\begin{abstract}
We consider the classical fluctuations of the gravitational constant
generated by bubbles in the inflationary universe. For extended inflation,
we demonstrate numerically how and how large fluctuations are produced
during bubble expansion. The amplitude of the fluctuations depends on the
Brans-Dicke parameter $\omega$: if $\omega$ is of the order of unity, the
amplitude becomes of the order of unity within one Hubble  expansion time;
if $\omega$ is large (say, $\omega=1000$), the growth rate of the
fluctuations is small, but it keeps growing without freezing during
inflation. We also discuss some astrophysical implications of our results.
\end{abstract}

\vskip 1cm
\begin{center}
To appear in {\it Physical Review D}
\end{center}
 
\newpage
\baselineskip = 18pt

Particle physics predicts that the universe experienced many phase 
transitions in its early history. If any of these phase transitions is
first order, the universe changes its phase from false vacuum to true
vacuum through the creation, expansion and collision of bubbles. Old
inflation \cite{oi} is based on a super-cooled first-order phase
transition. In the inflationary scenario, the universe expands
exponentially with time before the transition, thereby solving the
horizon, flatness, and monopole problems. However, it turns out that
this exponential expansion is too rapid to permit a transition from
false vacuum to true vacuum via percolation of true vacuum bubbles
\cite{oibad}.

Extended inflation \cite{ei} revived the idea of old inflation by using
the Brans-Dicke theory instead of the Einstein theory. The Brans-Dicke
field decelerates the expansion of the universe so that true vacuum
bubbles can coalesce, thus ending the phase transition that drives
inflation. This model provides an interesting hypothesis that the
large-scale structure of galaxy distribution is generated by primordial
bubbles \cite{la}. Many discussions about extended inflation have been 
made in past years. La {\it et al.} and Weinberg \cite{eibad} 
estimated
volume fraction of bubbles and found a difficulty: the constraint from
the isotropy of the cosmic microwave background requires the
Brans-Dicke parameter $\omega<25$, which contradicts the lower limit
$\omega>500$ from the time-delay experiments \cite{w500}. This problem 
is called a "big-bubble problem" because many big bubbles cause
large-scale inhomogeneity. Thus interest shifted to other extended
models based on general scalar-tensor theories \cite{get}-\cite{hei2}. 

In discussing density fluctuations in extended inflation or other extended
models, several authors have estimated volume fraction of bubbles as we
mentioned above, or quantum fluctuations of the gravitational constant 
(the Brans-Dicke field) \cite{QF1}. In this paper, we investigate the 
classical fluctuations of the gravitational constant generated during
bubble expansion. It was shown that, under the thin-wall approximation, the
Brans-Dicke field must be inhomogeneous inside a bubble from the
consistency of the junction conditions \cite{BubDyn}. However, it is 
not clear how and how large fluctuations are really generated. If these
fluctuations are not small, we should consider it for the constraint 
for the models, such as the structure formation process and the
measurements  of the gravitational constant.

There is an alternative scenario of inflation which is accompanied by
bubble nucleation: one-bubble inflation \cite{onebub1}. This model is
distinguishable from extended inflation or its generalized version:
the second slow-rollover inflation occurs inside a nucleated bubble, 
and our observable universe is entirely contained in one bubble. If we
assume the O(4)-symmetric bubble, the interior of the bubble can be a
homogeneous and isotropic open universe \cite{o4}; this model may account
for the  universe model with
$\Omega_0\mbox{\raisebox{-1.ex}{$\stackrel{\textstyle <}{\textstyle
\sim}$}}0.1$, which is supported by increasing observations. Although the
fluctuations of the gravitational constant may
not be relevant to this model, it is interesting to study the 
fluctuations inside a bubble in a similar way.

To begin with, we explain briefly why fluctuations of the gravitational
constant are generated in the inflationary universe even at a classical
level. Let us consider the system which is described by the action
\begin{equation}\label{action}
S = \int d^4x \sqrt{-g} \biggl[{1\over16\pi G}\Bigl\{\Phi {\cal R} -
{\omega(\Phi)\over\Phi}(\nabla\Phi)^2\Bigr\}
-U(\Phi)-{1\over2}(\nabla\psi)^2 - V(\psi)\biggr],
\end{equation}
where $\Phi$ is a Brans-Dicke-like scalar field which is normalized to 
be unity at the present epoch, and $\psi$ is a Higgs-like inflaton field.
The field equation for $\Phi$ is derived from (\ref{action}):
\begin{equation}\label{BDeq}
\kern1pt\vbox{\hrule height 1.2pt\hbox{\vrule width1.2pt\hskip 3pt
\vbox{\vskip 6pt}\hskip 3pt\vrule width 0.6pt}\hrule height
0.6pt}\kern1pt
\Phi = {1\over2\omega(\Phi)+3} \left[-8\pi
G\left\{(\nabla\psi)^2+V(\psi)\right\}
-\omega'(\Phi)(\nabla\Phi)^2+U'(\Phi)\right].
\end{equation}
Just looking at (\ref{BDeq}), we can understand that the inhomogeneity
of $\psi$ causes that of $\Phi$: even if $\Phi$ and $\psi$ are
homogeneous at the beginning of inflation, bubble nucleation, which is
realized by the inhomogeneization of $\psi$, makes $\Phi$
inhomogeneous inside a bubble. Because most bubbles become
superhorizon scale at the end of inflation, the fluctuations of the
gravitational constant, which trace these bubbles, cannot disappear 
soon after inflation.

We may note, in passing, that the inhomogeneization inside a bubble
occurs in generic systems. After a conformal transformation and the
introduction of a new scalar field $\phi$, many scalar-tensor theories 
have the Lagrangian form \cite{con}
\begin{equation}
S = \int d^4x \sqrt{-g} \biggl[{{\cal R}\over2\kappa^2}-\frac12(\nabla
\phi)^2-W(\phi) -\frac12e^{-\gamma\kappa\phi}(\nabla\psi)^2
-e^{-\beta\kappa\phi}V(\psi) \biggr],
\end{equation}
where $\beta$ and $\gamma$ are dimensionless coupling constants, and
$\kappa^2\equiv8\pi G$. The field equation for $\phi$ is written as
\begin{equation}
{\kern1pt\vbox{\hrule height 1.2pt\hbox{\vrule width1.2pt\hskip 3pt
\vbox{\vskip 6pt}\hskip 3pt\vrule width 0.6pt}\hrule height 0.6pt}\kern1pt}
\phi = -{\gamma\kappa\over2}e^{-\gamma\kappa\phi}(\nabla\psi)^2
-\beta\kappa e^{-\beta\kappa\phi}V(\psi) + W'(\phi).
\end{equation}
This equation also indicates that the inhomogeneity of $\psi$ causes
that of $\phi$. Thus we see that the inhomogeneity inside a bubble is 
a generic property in generalized Einstein theories or in multiple 
scalar field theories.

In what follows we numerically investigate how and how large fluctuations 
of $\Phi$ are really generated in the inflationary era. In order to solve 
the field equations, we must specify a gravitational theory, i.e., 
the functions of $\omega(\Phi)$ and $U(\Phi)$. Here we simply assume
$\omega(\Phi)=$ constant and $U(\Phi)=0$, which corresponds to extended
inflation. Although our assumption is just for simplicity, our results 
will be also applied to some of the models with variable $\omega$, 
so-called hyperextended inflation \cite{hei1,hei2}. Steinhardt and 
Accetta \cite{hei1} and Garcia-Bellido and Quiros \cite{hei2}
independently proposed similar scenarios: $\omega(\Phi)$ is small and
almost constant during inflation and diverges as $\Phi$ approaches
$\Phi_0(=1)$. As long as we discuss their models, the assumption of
constant $\omega$ during inflation is justified.

We assume a spherically symmetric spacetime:
\begin{equation}\label{sphere}
ds^2=-dt^2+A^2(t,r)dr^2+B^2(t,r)r^2(d\theta^2+\sin^2\theta d\varphi^2),
\end{equation}
As a potential of the inflaton field, we adopt the quartic potential 
which was used in \cite{SamHis}:
\begin{equation}\label{pot}
V(\psi) = (\psi-m)^2 \left\{3(\epsilon+1)\psi^2+2\epsilon m\psi
+\epsilon m^2 \right\}.
\end{equation}
Free parameters are $\epsilon$ and $m$: $\epsilon$ determines the shape
of the potential and the ratio of $m$ to the Planck mass $m_{{\rm Pl}}$
implies the strength of gravity. We depict some shapes of the 
potential in Fig. 1.

As an initial configuration of the inflaton field, $\psi(t=0,r)$, we
give a static bubble solution in a flat spacetime by numerically
solving the field equation. As for the Brans-Dicke field, we assume
$\Phi(t=0,r)$ to be homogeneous. We also suppose that
$\partial\psi/\partial t(t=0,r)=\partial\Phi/\partial t(t=0,r)=0$.
To solve the time-dependent field equations, we use a finite difference
method with 1000 meshes; the details of the method were shown in the
Appendix of \cite{topoinf}. The Hamiltonian constraint equation is
used for checking the numerical accuracy; through all the calculations
executed here, the errors are always less than a few  percent.

Our numerical results are summarized in Figs. 2$-$5. We normalize the
time scale and the spatial scale by $\chi^{-1}\equiv \{8\pi GV(0)/
3\Phi(0)\}^{-\frac12}$, which corresponds to the horizon scale at
the nucleation time. In Fig. 2 we show examples of the time-evolution of
$\psi$ and $\Phi$. We see that the fluctuations of $\Phi$ are really
generated as a bubble expands.

Here we define a quantity which corresponds to the amplitude of the
fluctuations as
\begin{equation}\label{delta}
\Delta(t) \equiv{\Phi(t,r_{{\rm max}})-\Phi(t,0)\over\Phi(t,r_{{\rm max}})},
\end{equation}
where $r_{{\rm max}}$ is the outer numerical boundary. We utilize 
$\Delta(t)$ to present the following results. In Fig. 3(a) we draw
$\Delta(t)$ for various $\omega$. As we can expect from (\ref{BDeq}),
$\Delta(t)$ at fixed $t$ is almost proportional to $1/\omega$: if
$\omega=1$, the amplitude becomes of the order of unity within one Hubble
expansion time; if $\omega=1000$, the growth rate of the fluctuations is
small, but $\Delta(t)$ goes on increasing  without freezing until one
bubble collides with another. 

Samuel and Hiscock \cite{SamHis} investigated Euclidean solutions of 
an O(4)-symmetric bubble and found their configurations for $m/m_{{\rm Pl}}$ of 
the order of unity are quite different from those of ordinary bubble
solutions. Here we pay attention to the behavior of $\Phi(t,r)$ for large
$m/m_{{\rm Pl}}$. Figure 4 shows the evolution of $\Phi(t,r)$ for $m/m_{{\rm
Pl}}=0.5$. The behavior of $\Phi(t,r)$ is quite different from that for small
$m$ [cf. Fig. 2 (b)]. In Fig. 4(b) we draw $\Delta(t)$ for $m/m_{{\rm
Pl}}=0.1,~0.2$, and $0.5$. As $m/m_{{\rm Pl}}$ becomes larger, the fluctuations
of $\Phi$ become larger. We can interpret our results as follows. In the case of
large $m/m_{{\rm Pl}}$, the viscosity term of $\partial\Phi/\partial t$ in the
right-hand side of (\ref{BDeq}) is dominant to spatial derivatives, and
therefore the homogeneization process is slowed.

We also investigate the dependence of the fluctuations of $\Phi$ on the
potential shape. We draw $\Delta(t)$ for several $\epsilon$ in Fig. 5. The
potential for $\epsilon=0.05$ corresponds to the case of a thin-wall
bubble, and $\epsilon=0.5$ to a thick-wall bubble. The temporal behavior
of $\Delta(t)$ for small $\epsilon$ (thin-wall bubble) is violent, but 
the eventual behavior depends little on $ \epsilon$.

So far we have numerically analyzed the evolution of a bubble in extended
inflation and see how and how large fluctuations of the Brans-Dicke field
are generated. In order to understand the astrophysical implications of 
our results, we also have to analyze their subsequent evolution after the
phase transition. Because most fluctuations are nonlinear and gravity
is not described by the Einstein theory, however, such analysis is not
easy: another simulation or some sophisticated approaches are needed. In
the rest of the paper, we offer discussion on the rough implications to
some inflationary models.

In the scenario of extended inflation or hyperextended inflation, all
bubbles are nucleated with the subhorizon size and cross outside the
horizon during inflation. After the phase transition, matter trace the
bubbles$-$we call such bubblelike distribution after inflation ``voids"$-$and
reenter again during the radiation-dominated era or the matter-dominated era.
Bubbles nucleated the earliest cross outside the horizon first, reenter last,
and become the largest voids at present. Although we do not know the exact
evolution of voids, we may suppose that homogeneization inside the void does
not occur effectively until the void reenters the horizon. Because
astrophysical-sized voids ($30-100$Mpc) reenters the horizon at
$z\approx10^4$ to $10^3$, the inhomogeneity of the gravitational constant
affects the evolution of voids during the radiation-dominated universe. In
relation to the big-bubble problem of extended inflation, Vadas \cite{vadas}
studied thermalization of superhorizon voids in the radiation-dominated
universe and found homogeneization can occur in a much shorter time than
previously thought. This result indicates that there remains a
possibility that the big-bubble problem is resolved; it is worth
studying further. In order to analyze the thermalization process more
precisely, it is important to take the inhomogeneity of the
gravitational constant into account.

Let us discuss the following question: is it possible to observe the
fluctuations of the gravitational constant at present? According to the linear
perturbation analysis in the matter-dominated Brans-Dicke universe
\cite{hirai}, the decay rate of the amplitude of $\Phi$ is approximately
proportional to (scale factor)$^{-{3\over4}}$ for $\omega\gg1$. This
indicates that, if $\delta\phi/\phi<1$ when a void enters the horizon
and the linear perturbation theory can be applied, the present
fluctuations of $\Phi$ inside astrophysical-sized voids are too small to
be observable. However, if the fluctuations are nonlinear at the time,
it may be possible that a large inhomogeneity still remains although the
amplitude cannot be estimated without further analysis. Small-scale
fluctuations caused by small bubbles are also interesting to study.
While Turner and Wilczek \cite{turner} pointed out the detectability of
the gravitational waves produced by bubble collisions in extended
inflation, small-scale bubbles could be a source of ``scalar-type"
gravitational waves which can be detected by a laser interferometer. 

Finally, we make some comments on one-bubble inflation. If we consider a single
scalar field in the Einstein theory, which is described by the O(4)-symmetric
bounce solution \cite{o4}, the interior of the bubble can be a homogeneous,
isotropic, and open universe. We may understand that the high symmetry allows
the inside spacetime to have a homogeneous and  open slicing by a miracle. On
the other hand, if we suppose two scalar fields or a modified Einstein
gravity, O(4) symmetry is lost and the inside must be inhomogeneous. Our
analysis gives a generic picture that in such a theory the fluctuations grow
during bubble expansion. Unless the amplitude of the fluctuations are
negligibly small at the end of inflation, the inhomogeneity can be observed
at present because it is still on a super-horizon scale. Recently, Linde and
Mezhlumian \cite{onebub2} proposed some models with two scalar fields, taking
into account the condition that the inhomogeneity inside a bubble should be
negligible. It may also be important to consider the classical fluctuations in
models with a modified Einstein gravity such as proposed in \cite{onebub3}.

The author would like to thank T. Hirai and A. Linde for enlightening comments.
Thanks are also due to W. Rozycki for correcting the manuscript. This work was
supported partially by the Grant-in-Aid for Scientific Research Fund of the
Ministry of Education, Science and Culture (No. 07740226), and by a Waseda
University Grant for Special Research Projects.


\vskip 1cm \noindent
\baselineskip = 18pt
{\Large{\bf Figure Captions}}
\vspace{.3 cm}

\noindent
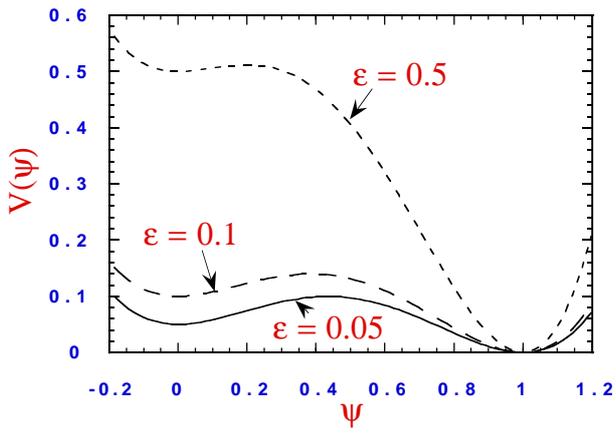
{\bf FIG. 1}. The potential of the inflaton field we used in our
analysis. We depict the shapes for $\epsilon=0.05, 0.1$, and $0.5$.

\vskip .5cm\noindent
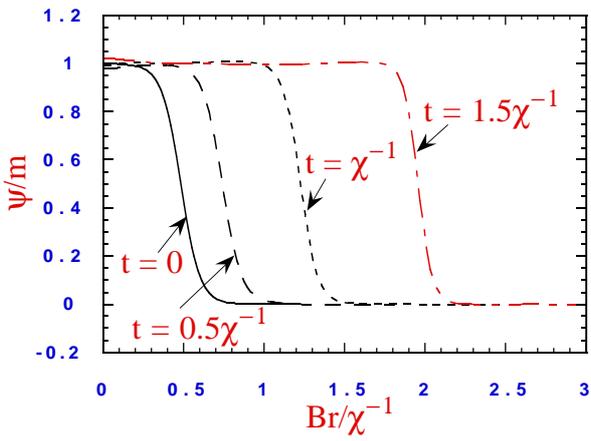
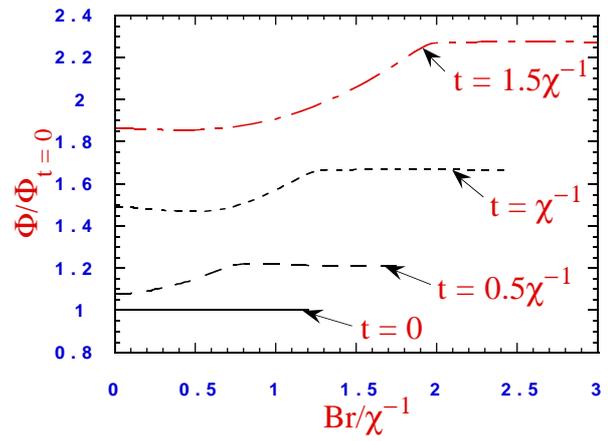
{\bf FIG. 2}. Examples of the evolution of (a) $\psi$ and (b) $\Phi$. 
We see how the fluctuations of $\Phi$ are generated as a bubble expands.

\vskip .5cm\noindent
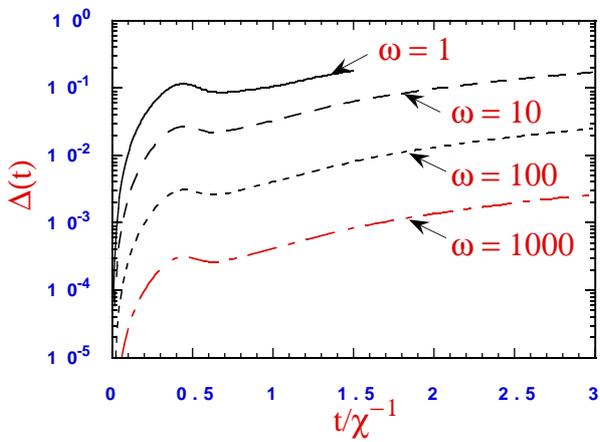
{\bf FIG. 3}. Dependence of the fluctuations of $\Phi$ on $\omega$. We
draw the amplitude $\Delta(t)$ for $\omega=1,~10,~100$, and $1000$.
We set $\epsilon=0.1$ and $m=0.1$. $\Delta(t)$ at fixed $t$ is almost
proportional to $1/\omega$, but it goes on increasing even for large
$\omega$. 

\vskip .5cm\noindent
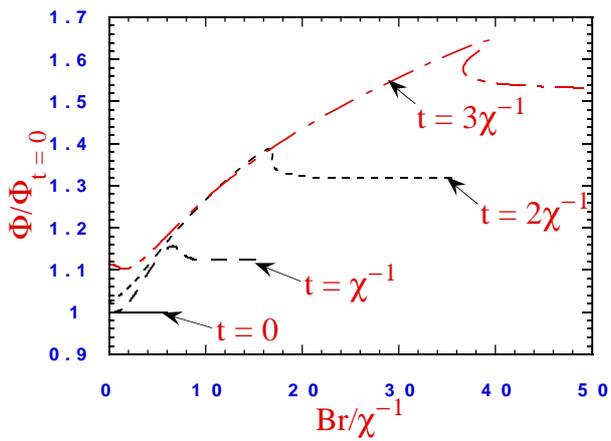
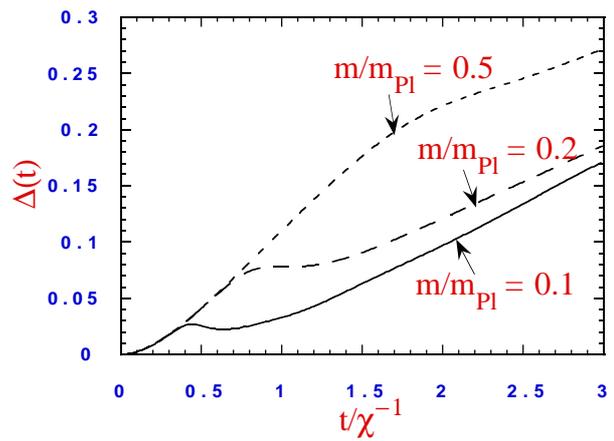
{\bf FIG. 4}. Dependence of the fluctuations of $\Phi$ on $m/m_{Pl}$. 
We set $\omega=10$ and $\epsilon=0.1$. In (a) we show the evolution of
$\Phi(t,r)$ for $m/m_{Pl}=0.5$. The behavior of $\Phi(t,r)$ is quite
different from that for small $m$. In (b) we draw $\Delta(t)$ for
$m/m_{Pl}=0.1,~0.2$, and $0.5$. As $m/m_{Pl}$ is larger, the
fluctuations of $\Phi$ become larger.

\vskip .5cm\noindent
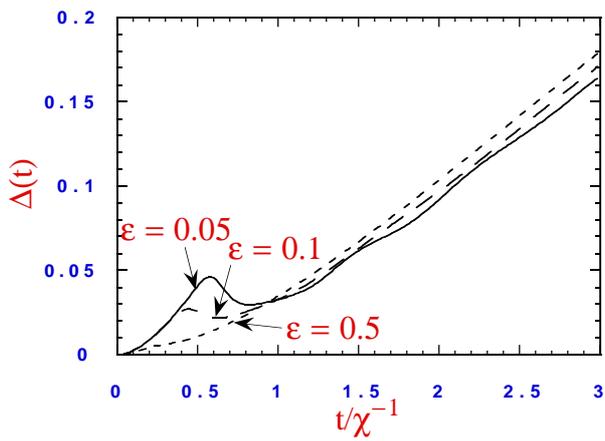
{\bf FIG. 5}. Dependence of the fluctuations of $\Phi$ on the potential
shape. We draw $\Delta(t)$ for $\epsilon=0.05,~0.1$, and $0.5$. We
set $\omega=10$ and $m=0.1$. Although the temporal behaviors of
$\Phi(t,r)$ are quite different, the eventual behavior does not depend
on the potential shape.


\begin{thebibliography}{99}
\baselineskip = 18pt

\bibitem{oi}A.H. Guth, Phys. Rev. D {\bf 23}, 347 (1981);\\ 
K. Sato, Mon. Not. R. Astron. Soc. {\bf 195}, 467 (1981).
\bibitem{oibad}A.H. Guth and E.J. Weinberg, Phys. Rev. D {\bf 23}, 321
(1981); Nucl. Phys. {\bf B212}, 321 (1983).
\bibitem{ei}D. La and P.J. Steinhardt, Phys. Rev. Lett. {\bf 62}, 376
(1989); Phys. Lett. B {\bf 220}, 375 (1989).
\bibitem{la}D. La, Phys. Lett. B {\bf 265}, 232 (1991).
\bibitem{eibad}D. La, P.J. Steinhardt, and E.W. Bertschinger, Phys. 
Lett. B {\bf 231}, 231 (1989);\\ 
E.J. Weinberg, Phys. Rev. D {\bf 40}, 3950 (1989).
\bibitem{w500}R.D. Reasenberg {\it et al.}, Astrophys. J. {\bf 234}, L219
(1979).
\bibitem{get}F.S. Accetta and J.J. Trester, Phys. Rev. D {\bf 39}, 2854 (1989);
\\R. Holman, E.W. Kolb, and Y. Wang, Phys. Rev. Lett. {\bf 65}, 17 (1990);\\
J.D. Barrow and K. Maeda, Nucl. Phys. {\bf B341}, 294 (1990).
\bibitem{hei1}P.J. Steinhardt and F.S. Accetta, Phys. Rev. Lett. {\bf 64}, 2740 (1990).
\bibitem{hei2}J. Garcia-Bellido and M. Quiros, Phys. Lett. B {\bf 243}, 45 
(1990).
\bibitem{QF1}S. Nakamura and A. Hosoya, Prog. Theor. Phys. {\bf 87}, 
401 (1992);\\
J. Garcia-Bellido, A. Linde, and D. Linde, Phys. Rev. D {\bf 50}, 730 (1994).
\bibitem{BubDyn}D.S. Goldwirth and H.W. Zaglauer, Phys. Rev. Lett.
{\bf 26}, 3639 (1991); 
N. Sakai and K. Maeda, Phys. Rev. D {\bf 48}, 5570 (1993); Prog. Theor. Phys.
{\bf 90}, 1001 (1993).
\bibitem{onebub1}M. Sasaki, T. Tanaka, K. Yamamoto, and J. Yokoyama,
Phys. Lett. B {\bf 317}, 510 (1993);\\
M. Bucher, A.S. Goldhaber, and N. Turok, Phys. Rev. D {\bf 52}, 3314 (1995).
\bibitem{o4}S. Coleman and F.De Luccia, Phys. Rev. D {\bf 21}, 3305 (1980).
\bibitem{con}K. Maeda, Phys. Rev. D {\bf 39}, 39 (1989);\\
A.L. Berkin and K. Maeda, {\it ibid.} {\bf 44}, 1691 (1991).
\bibitem{SamHis}D.A. Samuel and W.A. Hiscock, Phys. Rev. D {\bf 44}, 3052
(1991).
\bibitem{topoinf}N. Sakai, H. Shinkai, T. Tachizawa and K. Maeda,
Phys. Rev. D {\bf 53}, 655 (1996).
\bibitem{vadas}S.L. Vadas, Phys. Rev. D {\bf 48}, 4562 (1993).
\bibitem{hirai}T. Hirai and K. Maeda, Astrophys. J. {\bf 431}, 6 (1994);\\
H. Nariai, Prog. Theor. Phys. {\bf 42}, 544 (1969).
\bibitem{turner}M.S. Turner and F. Wilczek, Phys. Rev. Lett. {\bf 65}, 3080
(1990).
\bibitem{onebub2}A. Linde, Phys. Lett. B {\bf 351}, 99 (1995);\\
A. Linde and A. Mezhlumian, Phys. Rev. D {\bf 52}, 6789 (1995).
\bibitem{onebub3}L. Amendola, C. Baccigalupi, and F. Occhionero, Report No.
astro-ph/9504097.
\end{thebibliography}
\end{document}